\documentclass[12p]{article}
\usepackage{eso-pic} 
\usepackage{latexsym}
\usepackage{color}
\usepackage{amsmath}
\usepackage{epsfig}
\usepackage{hyperref}
 \usepackage{subfigure}
 \usepackage{dcolumn}
   \usepackage{threeparttable}
     \newcommand{\thickhline}{\noalign{\hrule height 0.8pt}}
\newcommand{\ortala}[1]{\begin{center}#1\end{center}}

\newcommand{\sandd}[1]{\left\langle #1\right\rangle}
\newcommand{\sanddr}[1]{\left\langle\left\langle #1\right\rangle\right\rangle_r}
\newcommand{\integ}[3]{{{\underset{#1 }{\overset{#2}{\displaystyle\int}}}#3}}

\newcommand{\summ}[3]{{{\underset{#1 }{\overset{#2}{\displaystyle\sum}}}#3}}
\newcommand{\prodd}[3]{{{\underset{#1
}{\overset{#2}{\displaystyle\prod}}}#3}}

\newcommand{\re}[1]{(\ref{#1})}

\newcommand{\eq}[2]{\begin{equation}\label{#1}  #2\end{equation}}

\newcommand{\paran}[1]{\left(#1\right)}

\newcommand{\sch}[1]{Schrodinger}

\linethickness{0.6pt}

\begin{document}

\ortala{\textbf{An improved effective field theory formulation of
spin-1/2 Ising systems with arbitrary coordination number $z$}}

\ortala{\textbf{\"Umit Ak\i nc\i \footnote{umit.akinci@deu.edu.tr}}}

\ortala{\textit{Department of Physics, Dokuz Eyl\"ul University,
TR-35160 Izmir, Turkey}}

\section{Abstract}

An improved  unified formulation based on the effective field theory
is introduced for a spin-1/2 Ising model with nearest neighbor
interactions with arbitrary coordination number $z$. Present
formulation is capable of calculating all the multi-spin
correlations systematically in a representative manner , as well as
its single site counterparts in the system and gives results for the
critical temperature of the system much better than those of the
other works in the literature. The formulation can be easily applied
to various extensions of $s$-1/2 Ising models, as long as the system
contains only the nearest neighbor interactions as spin-spin
interactions. Keywords: \textbf{Ising model ; Effective field theory
;Correlation functions }

\section{Introduction}\label{intro}
Ising model \cite{ref1} has been one of the most extensively studied
models in statistical mechanics and condensed matter physics for a
long time. The reason is due to the fact that this simple many-body
system suits well for investigation of thermal and magnetic
properties of various physical systems. From the theoretical point
of view, this model has been handled by a variety of approximation
techniques and each of them were developed for a demand of obtaining
better results, especially for determination of critical temperature
of the system as accurate as the known exact results. Bethe-Peierls approximation (BPA) \cite{refBPA}, Bethe lattice approximation (BLA) \cite{refBLA}
the effective field theory (EFT) \cite{refEFT} the correlated effective field theory (CEFT) \cite{refCEFT} the
cluster variation method (CVM) \cite{refCVM} the series expansion methods (SE) \cite{refSE} and
expanded Bethe-Peierls approximation (EBPA) \cite{refEBPA} are among the most widely used approximations for the s-1/2 Ising systems.
Among these methods, EFT in which the system is treated
as a finite cluster is regarded as one of the most powerful methods
which gives rather accurate results than those of conventional mean
field theory which considers a many-body system as a single particle
system.


In a typical EFT method \cite{refEFTK1} , one writes the spin
interactions on a finite cluster explicitly by choosing an
appropriate Hamiltonian. Contributions coming from the outside of
the defined cluster are represented by an effective field, and
determination of this effective field plays a key role for solving
the problem. Another alternative choice is to work on a larger
cluster and to define the interactions  only within this larger
finite cluster \cite{ref5},\cite{ref6}. In EFT methods, calculations
are often carried on by starting with exact spin identities
\cite{ref3},\cite{ref4}.  These identities include the thermal
average of some certain functions which depend on the spin variables
in the cluster. Hence, in addition to the single spin cluster
identities there are several works based on two or more spin cluster
identities in the literature \cite{ref5},\cite{ref6}. Whatever form
of identities used, in order to perform the thermal averages,
differential operator \cite{refEFT} and integral operator techniques
\cite{ref8},\cite{ref9} are often used. However, the accuracy and
responsibility of the results obtained by these EFT methods costs
some complicated mathematical difficulties. Namely, it is widely
believed that if one tries to treat exactly all the spin-spin
correlations emerging on the expansion of the spin identities then
the problem becomes mathematically intractable. In order to overcome
these difficulties, one often has to make an approximation on the
problem of evaluating the correlation functions which means a less
accuracy of the results in comparison with the exact ones. Hence, it
is no matter that whatever the technique is used, the most of the
studies in the literature refer to a decoupling approximation (DA)
which neglects the multi-spin correlations \cite{refEFT} and gives
the results identical to those of Zernike approximation
\cite{refZER}. However there are small number of works try to handle
these correlations \cite{refCEFT},\cite{refEFTK1}. Although they
have different formulations, these two works gives results which are
identical to those of BPA for the critical temperatures.

Recently, we have introduced an improved formulation, based on
differential operator technique for investigation of the thermal and
magnetic properties of a 3D random field Ising model \cite{ref10}.
Our formulation improves the results of the conventional EFT methods
by taking into account the multi-spin correlations. In this paper,
we wish to extend the method given in Ref. \cite{ref10} for a
general Ising s-1/2 system with nearest neighbor interactions by
proposing a unified formulation which systematically includes the
multi-spin correlation effects for a given arbitrary coordination
number $z$. The general formulation being presented here can be
easily adopted to more complicated systems such as the models with
transverse field. For this purpose, we organized the paper as
follows: In Sec. \ref{formulation} we explicitly present the
formulations. The numerical results and discussions are summarized
in Sec. \ref{results} with application of formulation  to pure
system and bond diluted system and finally Sec. \ref{conclusion}
contains our conclusions.

\section{General formulation}\label{formulation}
The Hamiltonian describing our model is
\begin{equation}\label{eq1}
\mathcal{H}=-\summ{<i,j>}{}{J_{ij}s_is_j-\summ{i}{}{H_is_i}}
\end{equation}  Here $s_i$ denotes the $z$ component of the spin variable and it takes the values
$s_i=\pm 1$,   $J_{ij}=J_{ji}>0$ is the ferromagnetic exchange
interaction between spins $i$ and $j$,
and $H_i$ is the external longitudinal magnetic field on a site $i$. The first
summation in Eq. \re{eq1} is over the nearest-neighbor pairs of
spins and the other summation is over the all lattice sites. The quantities
$J_{ij}$, and $H_i$ may be given with certain distributions or
they can have the same values for all pairs of spins/sites, i.e $J_{ij}=J,
H_i=H$.

As our model system, we consider a regular lattice which has $N$ identical spins arranged with coordination number z.
We define a cluster on the lattice which consists of a central spin labeled $s_0$, and $z$ perimeter spins ($s_\delta,\delta=1,2,\ldots z$) being
the nearest-neighbors of the central spin. The nearest-neighbor spins are in an effective field produced by the outer spins,
which can be determined by the condition that the thermal and configurational average of the central spin is equal to that of its nearest-neighbor spins.

According to the Callen identity \cite{ref3}, averages of
the spin variables in the cluster are given by
\begin{equation}\label{eq2}
\sanddr{\{f_i\}s_i}=\sanddr{\{f_i\}\tanh{\paran{\beta E_i}}}
\end{equation}
where $\beta = 1/k_BT$, $k_B$ is Boltzmann constant and $T$ is the temperature,  $E_i$ is the part of the Hamiltonian in
Eq. \re{eq1} which includes all contributions associated with the site $i$ and $\{f_{i}\}$ is an arbitrary function which is
independent of the site $i$. The inner average bracket (which has no subscript)  stands for thermal average and the outer one (which has subscript $r$) is
for configurational average which is necessary for including the effect of random bond and random field distributions.

At this stage, Eq. \re{eq2} is exact but it is not possible to perform the averages without making an approximation due to
the large number of degrees of freedom of the system. One coarse approximation is well known mean field approximation.
In this approximation (for the system homogenous distributed bonds, i.e. $J_{ij}=J$ for all $i,j$ and zero magnetic field) the spin $s_i$ interacts with the field $Jz\sandd{s}$ where $\sandd{s}=\sandd{s_i},i=0,1,\ldots ,z$. Thus,
in this approximation magnetization is given by
\begin{equation}\label{eq3}
\sandd{s}=\tanh{\paran{\beta Jz \sandd{s}}}
\end{equation}
On the other hand within the EFT, average of the central spin is
obtained from Eq. \re{eq2} with
$E_0=-\summ{\delta=1}{z}{J_{0\delta}s_\delta}-H_0$ as \eq{eq4}{
\sanddr{\{f_0\}s_0}=\sanddr{\{f_0\}\tanh{\paran{\beta
\summ{\delta=1}{z}{J_{0\delta} s_\delta}+\beta H_0}}}.} By choosing
$\{f_0\}=1$, we can get the thermal and configurational average of a
central spin from Eq. (\ref{eq4}). In a similar manner, perimeter
spin in the cluster, interacts with central spin and with an
effective field which is produced by the $(z-1)$ spins outside of
the cluster. Hence, the average of the perimeter spin can be
obtained again from Eq. \re{eq2}, but this time with
$E_\delta=-J_{0\delta}s_0-(z-1)h-H_\delta$.
\begin{equation}\label{eq5}
\sanddr{\{f_\delta\}s_\delta}=\sanddr{\{f_\delta\}\tanh{\paran{\beta J_{0\delta} s_0+\beta (z-1)h+\beta H_\delta}}}
\end{equation}
where $h$ is the effective field per spin. By choosing $\{f_\delta\}=1$, we get the average of the perimeter spin from Eq. (\ref{eq5}) .
Since all the sites are equivalent in the model, the value of the effective field $h$ for a fixed temperature and certain distributions can be determined from the relation
\begin{equation}\label{eq6}
\sanddr{s_0}=\sanddr{s_\delta}
\end{equation}
Hence, for a given $(\beta,h)$ pair with a certain bond and longitudinal magnetic field distribution,
we can obtain magnetization as a function of the temperature. Since the effective field is very small in the vicinity of critical
temperature then the critical temperature can be obtained by letting $h\rightarrow 0$ in Eq. \re{eq6}
and solving it for $\beta=\beta_c$ where $\beta_c = 1/k_BT_c$, $T_c$ is the critical temperature.

In order to evaluate right hand sides of Eqs. \re{eq4} and \re{eq5}
we can apply the differential operator technique \cite{refEFT}
\eq{eq10}{\exp{\paran{a\nabla}}f\paran{x}=f\paran{x+a}}
where $\nabla=\partial/\partial x$ is the differential operator and $a$ is any constant. With the help of Eq. \re{eq10}
we
get the following equations
\begin{equation}\label{eq7}
\sanddr{\{f_0\}s_0}=\sanddr{\{f_0\}\prodd{\delta=1}{z}{\left[\cosh{\paran{J_{0\delta}\nabla}}+s_\delta\sinh{\paran{J_{0\delta}\nabla}}\right]}}f(x)|_{x=0}
\end{equation}
\begin{equation}\label{eq8}
\sanddr{\{f_\delta\}s_\delta}=\sanddr{\{f_\delta\}\left[\cosh{\paran{J_{0\delta}\nabla}}+s_0\sinh{\paran{J_{0\delta}\nabla}}\right]}f(x+(z-1)h)|_{x=0}
\end{equation}
where
\begin{equation}\label{eq9}
f(x)=\integ{}{}{dH_i P_H(H_i)g(x,H_i)},\quad
g(x,H_i)=\tanh{\left[\beta\paran{ x+H_i}\right]}.
\end{equation}  Here $P_H(H_i)$ is the probability distribution function
of  longitudinal magnetic field.

By expanding the right hand side of  Eq. \re{eq7} with $\{f_0\}=1$
we can obtain the thermal and configurational average of the central
spin. This expansion which is superior to conventional MFA takes
into account the self-spin correlation identity which is given by
\eq{eq10ek}{s_i^{2n}=1, s_i^{2n-1}=s_i} for all $i$ where $n$ is
positive integer and the expansion produces the multi-spin
correlations among the perimeter spins for s-1/2 Ising system. In
order to avoid dealing with the mathematical difficulties
originating from these multi-spin correlations, an approximation
which is widely used in the literature, i.e. DA \cite{refACTA}can be
used according to
\begin{equation}\label{eq11}
\sanddr{s_is_j\ldots s_k}=\sanddr{s_i}\sanddr{s_j}\ldots \sanddr{s_k}.
\end{equation}
However, the multi-spin correlations can be evaluated by using an appropriate $\{f_\delta\}$ variable and expanding Eq. \re{eq8}.
In other words, we can calculate any correlation  which includes only perimeter spins from Eq. \re{eq8} by using an
appropriate $\{f_\delta\}$. Moreover, if we expand the right hand side of Eq. \re{eq8} with a selected  $\{f_\delta\}$
some other correlations appear which are represented in terms of the central spin and perimeter spins in the cluster. Similarly, these correlations
can also be calculated by using Eq. \re{eq7} and by selecting a suitable $\{f_0\}$. As a result of this process, we obtain a set of linear
equations in which all correlations are treated as unknowns. Since this approximation takes into account
the multi-spin correlations, it gives better results than those obtained by EFT with DA.

Now, in order to obtain a system of linear equations with arbitrary coordination number $z$, let us write Eq. \re{eq7} as,
\eq{eq12}{\sanddr{\{f_0\}s_0}=\sanddr{\{f_0\}\summ{n=0}{z}{\left[\paran{\begin{array}{c}z\\n\end{array}}
\prodd{\delta=1}{n}{s_\delta\sinh{\paran{J_{0\delta}\nabla}}   } \prodd{\delta=n+1}{z}{}\cosh{\paran{J_{0\delta}\nabla}}\right]}}f(x)|_{x=0}}
While writing Eq. \re{eq7} as Eq. \re{eq12} we use an approximation that all $n$ multi-site correlations among the perimeter spins are equal to each other and we represent them
with $\sanddr{s_\delta^{(n)}}=\sanddr{s_1s_2\ldots s_n}$. In other words we represent $\paran{\begin{array}{c}z\\n\end{array}}$ possible $n$-site perimeter spin correlations with
one correlation. For example for $z=3,n=2$, the correlations
$$
\sanddr{s_1s_2}=\sanddr{s_2s_3}=\sanddr{s_1s_3}
$$ are identical and represented as $s_\delta^{(2)}=s_1s_2$. The same approximation holds also for multi-site correlations which
include central site $s_0$ i.e. $\sanddr{s_0s_\delta^{(n-1)}}=\sanddr{s_0s_1\ldots s_{n-1}}$ represents all $n$-site correlations which include the central site.

Now, since all the terms except $s_\delta$ are independent of the thermal average then we can write Eq. \re{eq12} in the form:
\eq{eq13}{\sanddr{\{f_0\}s_0}= \summ{n=0}{z}{}A_n\sanddr{\{f_0\}s^{(n)}_\delta}}
where $s_\delta^{(n)}=s_1s_2\ldots s_n$ and the coefficient $A_n$ is given by
\eq{eq14}{
A_n=\paran{\begin{array}{c}z\\n\end{array}}\sandd{\left[
\prodd{\delta=1}{n}{\sinh{\paran{J_{0\delta}\nabla}}   } \prodd{\delta=n+1}{z}{}\cosh{\paran{J_{0\delta}\nabla}}\right]}_r f(x)|_{x=0}
} The configurational average can be taken by using any given bond distribution probability function $P_J\paran{J_{ij}}$
\eq{eq15}{
A_n=\paran{\begin{array}{c}z\\n\end{array}}\left[\integ{}{}{}\prodd{\delta=1}{z}{}\left[dJ_{0\delta}P_J\paran{J_{0\delta }}\right]
\prodd{\delta=1}{n}{}\left[\sinh{\paran{J_{0\delta}\nabla}}\right] \prodd{\delta=n+1}{z}{}\left[\cosh{\paran{J_{0\delta}\nabla}}\right]\right]f(x)|_{x=0}
}

The derivation of the multi-site correlations by adopting a suitable $\{f_0\}$ in Eq. \re{eq13} requires the determination of the identity $\sanddr{\{f_0\}s^{(n)}_\delta}$ which lies on the right-hand side of Eq. \re{eq13}. Let us call choosing $\{f_0\}=s_k$ in this term as applying
$s_k$ to $\sanddr{s_\delta^{(n)}}$. Applying $s_k$ to $\sanddr{s_\delta^{(n)}}$ will yield another correlation. Resulting correlation may be $\sanddr{s_\delta^{(n-1)}}$
or $\sanddr{s_\delta^{(n+1)}}$ depending on the relation between $k$ and $n$. For determining the resulting correlation we use Eq. \re{eq10ek} and it will be given by
\eq{eq16}{
\sanddr{s_k s_\delta^{(n)}} =\left\{\begin{array}{lcl}\sanddr{s_\delta^{(n+1)}}&,& \quad k>n\\
\sanddr{s_\delta^{(n-1)}}&,& k\le n\quad \\
\end{array}\right.
}

According to Eq. \re{eq16}, if we select $\{f_0\}=s_1$ in Eq. \re{eq13} then we get
\eq{eq17}{\sanddr{s_0s_1}=A_1+\paran{A_0+A_2}\sanddr{s_1}+\summ{n=3}{z}{A_n \sanddr{s_\delta^{(n-1)}}}}
Similarly, if we apply  $s_2$ to Eq. \re{eq17} we obtain
\eq{eq18}{\sanddr{s_0s_1s_2}=\paran{A_1+A_3}\sanddr{s_1}+\paran{A_0+A_2+A_4}\sanddr{s_1s_2}+\summ{n=5}{z}{A_n \sanddr{s_\delta^{(n-2)}}}}
After successive operations, we can get a general expression for the correlation $\sanddr{s_0s_1s_2\ldots s_k}$ as follows
\eq{eq19}{\sanddr{s_0s_\delta^{(k)}}=\sanddr{s_\delta^{(k-1)}}\summ{n=0}{k-1}{A_{2n+1}}+\sanddr{s_\delta^{(k)}}
\summ{n=0}{k}{A_{2n}}+\summ{n=2k+1}{z}{A_n\sanddr{s_\delta^{(n-k)}}}}

Multi-site correlations which appear on the right hand side of Eq. \re{eq19}  can be obtained from Eq. \re{eq8}. Writing Eq. \re{eq8} for $\delta=1$
and choosing $\{f_1\}=s_2$ in this expression will yield

\eq{eq20}{
\sanddr{s_1s_2}=B_1 \sanddr{s_1}+B_2\sanddr{s_0s_1}
} where
\eq{eq21}{
\begin{array}{lcl}
B_1&=&\integ{}{}{dJ_{01} P_J\paran{J_{01}}\cosh{\paran{J_{01}\nabla}}}f(x+\gamma)|_{x=0}\\
B_2&=&\integ{}{}{dJ_{01} P_J\paran{J_{01}}\sinh{\paran{J_{01}\nabla}}}f(x+\gamma)|_{x=0}
\end{array}
}
Applying $s_3$ to Eq. \re{eq20} will yield the three site correlation which include only perimeter spins in the cluster:
\eq{eq22}{
\sanddr{s_1s_2s_3}=B_1 \sanddr{s_1s_2}+B_2\sanddr{s_0s_1s_2}
}

By successive derivations we find
\eq{eq23}{\sanddr{s_\delta^{(k)}}=B_1\sanddr{s_\delta^{(k-1)}}+B_2\sanddr{s_0s_\delta^{(k-1)}}}
which is the correlation expression for $k$ perimeter spins. One other possible choice for deriving correlations which include  central site is to apply $s_0$ in Eq. \re{eq23}\cite{refEFTK1}. As seen in Ref. \cite{refEFTK1} this procedure will produce the results of BPA.
By taking into account Eq. \re{eq10ek} we can write these correlations as
\eq{eq24}{\sanddr{s_0s_\delta^{(k)}}=B_1\sanddr{s_0s_\delta^{(k-1)}}+B_2\sanddr{s_\delta^{(k-1)}}}

Thus we can get a system of linear equations by labeling the correlations;
\eq{eq25}{\begin{array}{lcl}
x_{k}&=&\sanddr{s_\delta^{(k)}}, \quad k=0,1,\ldots,z; \quad x_0=1\\
x_{z+k+1}&=&\sanddr{s_0s_\delta^{(k)}}, \quad k=0,1,2,\ldots,z
\end{array}
}

By rewriting Eq. \re{eq19} according to Eq. \re{eq25} we get
\eq{eq26}{x_{z+k+1}=x_{k-1}T_{k-1}+x_{k}C_k+\summ{n=1}{z-2k }{A_{2k+n}x_{k+n}},\quad k=0,1,\ldots,z}
where
\eq{eq27}{C_k=\summ{n=0}{k}{A_{2n}}, \quad T_{k}=\summ{n=0}{k}{A_{2n+1}}.}
Similarly for Eq. \re{eq23} with Eq. \re{eq25} we have
\eq{eq28}{x_{k}=B_1x_{k-1}+B_2x_{z+k},\quad k=1,2,\ldots,z}
and for Eq. \re{eq24} with Eq. \re{eq25}
\eq{eq29}{x_{z+k+1}=B_1x_{z+k}+B_2x_{k-1},\quad k=1,2,\ldots,z}
Note that $x_{-1}$ has to be regarded as $0$ in Eq. \re{eq26}.

Finally, we construct the system of linear equations based on Eqs. \re{eq26} and \re{eq28}  with the coefficients given
in Eqs. \re{eq15},\re{eq21} and \re{eq27}. By solving this system with the condition
Eq. \re{eq6} (i.e $x_1=x_{z+1}$) we can get all representative correlations defined on the cluster and hence, we
obtain better results than DA for the critical temperatures of Ising systems with coordination number $z$.

Another alternative choice for constructing linear equation system is forming it from Eqs. \re{eq26} for only $k=0$, \re{eq28} and \re{eq29}.
The results obtained by forming equation system in this way corresponds to results in \cite{refEFTK1}. These are slightly different from results
of DA. Instead of this we form our equation system from Eqs. \re{eq26} and \re{eq28}. This will give better results than DA and results in  \cite{refEFTK1}
as seen in Section \ref{results}. This may be due to the fact that the equation system which formed from  Eqs. \re{eq26} and \re{eq28} includes smaller number of terms which have effective field than the system formed from Eqs. \re{eq26} for only $k=0$, \re{eq28} and \re{eq29}.

For completeness of the work, let us make discussions on the
DA. The equation of state corresponding to
DA can be obtained from Eq. \re{eq26} for
$k=0$ and  by letting $x_j=m^j$ as part of Eq. \re{eq11} where $m$
stands for $\sanddr{s_0}$. \eq{eq30}{A_0+\summ{n=1}{z}{A_{n}m^n}-m=0}

Solving \re{eq30} with coefficients given in Eq. \re{eq15} will give
the magnetization within the DA. The critical temperatures can be
obtained by letting $m\rightarrow 0$ in \re{eq30} and solving it for
$\beta=\beta_c$\cite{refACTA}.

\section{Results and Discussion}\label{results}

Since the aim of this paper is to introduce an improved formulation within the EFT which is capable of calculating the multi-site correlations for any coordination number $z$, we give
only few results in comparison with DA and EFT results. Hence, we wish to present the results of  the formulation presented here
only for a pure system with zero magnetic filed and for a bond diluted system (with bearing in mind that the full discussion of this problem is out of the scope of this paper) in comparison with DA and EFT results. The results of random field Ising model on three dimensional lattices can be found in our earlier work \cite{ref10}.

\subsection{Pure System with Zero Magnetic Field}

This is the simplest s-1/2 Ising system. The distribution functions are delta functions as $P_J\paran{J_{ij}}=\delta\paran{J_{ij}-J}$ for all $i,j$ pairs of spins and
$P_H\paran{H_i}=\delta\paran{H_i-H}$ where $H=0$ for all $i$. Under these distributions, the coefficients of our linear equation system which are given in Eqs.
\re{eq15} and \re{eq21} gets the form

\eq{eq31}{
\begin{array}{lcl}
A_n&=&\paran{\begin{array}{c}z\\n\end{array}} \cosh^{z-n}{\paran{J\nabla}}\sinh^{n}{\paran{J\nabla}}f(x)|_{x=0}\\
B_1&=&\cosh{\paran{J\nabla}}f(x+\gamma)|_{x=0}\\
B_2&=&\sinh{\paran{J\nabla}}f(x+\gamma)|_{x=0}
\end{array}
}

In Eq. \re{eq31}, if we write the hyperbolic functions $\cosh{\paran{J\nabla}}$ and $\sinh{\paran{J\nabla}}$ in terms of exponential
expressions $\exp{\paran{J\nabla}}$ and $\exp{\paran{-J\nabla}}$, and use the binomial expansion with
the differential operator technique in Eq. \re{eq10} then we can obtain more suitable forms of the coefficients $A_n,B_1,B_2$ as

\eq{eq32}{\begin{array}{lcl}
A_n&=&\frac{1}{2^z}\paran{\begin{array}{c}z\\n\end{array}}\summ{r=0}{z-n}{}\summ{s=0}{n}{\paran{\begin{array}{c}z-n\\r\end{array}}
\paran{\begin{array}{c}n\\s\end{array}}(-1)^s f[(z-2r-2s)J]}\\
B_1&=&\frac{1}{2}\left[f(J+\gamma)+f(-J+\gamma)\right]\\
B_2&=&\frac{1}{2}\left[f(J+\gamma)-f(-J+\gamma)\right]
\end{array}}

By solving the system of linear equations defined by Eqs. (\ref{eq26}) and (\ref{eq28}) for an arbitrary coordination number $z$ with
the coefficients given in Eqs. (\ref{eq27}) and (\ref{eq32})  we get the temperature dependence of the whole representative correlations
defined in the system, as well as the critical temperature as a function of $z$ by
taking the limit $\gamma\rightarrow 0$ in the coefficients.

In Table \ref{table1}, we compare the critical temperature
$k_{B}T_{c}/J$ of the system obtained within the present work with
those of the other methods in the literature for various
coordination numbers. As seen in Table  \ref{table1}, our
formulation gives the best approximated values for the numerical
values of the critical temperatures when compared with MFA, EFT
\cite{refEFTK1} and DA \cite{refACTA}, which originates from the
consideration of the multi-spin correlations in the system. The
exact results listed in Table \ref{table1} are the results of high
temperature series expansion method  \cite{refexact} which has been
regarded as the exact results in the literature except the case
$z=4$ which has analytical result \cite{refexactz4}.

The improvement of this formulation is not limited with the better critical temperatures when compared with other approximations.
The computability of correlations allows us to determine some of the thermodynamic functions of the system, such as the internal energy and the
specific heat.

\begin{figure}[h]\begin{center}
\epsfig{file=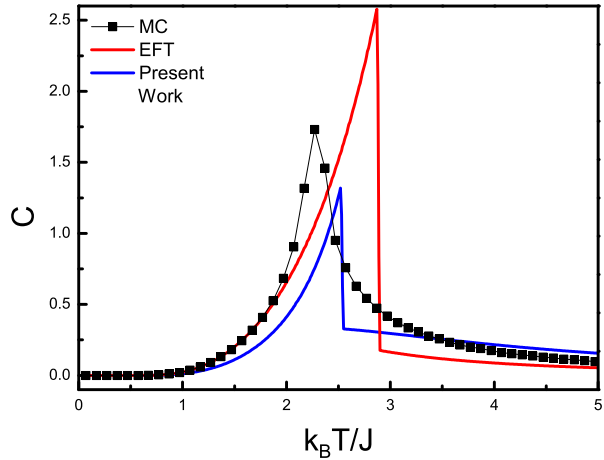, angle=0,width=8cm}
\end{center}
\caption{Variation of specific heat of square lattice with
temperature for pure system under zero magnetic field. While the
dotted curve shows the MC results, the red curve is EFT result and
the blue one is our formulation's result. }\label{fig3}\end{figure}

In Fig.  \ref{fig3},  we can see  the temperature dependence of the specific heat curves which are obtained from EFT and our formulation in comparison with Monte Carlo (MC) simulation for the square lattice ($z=4$). In MC simulation $100\times 100$ lattice and standard Metropolis Algorithm was used. Since MC simulation gives behavior for specific heat close to real one, we can conclude from Fig. \ref{fig3} that our formalism gives more accurate behavior of the specific heat than EFT. This means that formulation presented here
handles two site correlations more accurate than EFT and this results in a more accurate behavior of internal energy with temperature, as well as the behavior of specific heat with temperature.

The temperature dependencies of magnetization,  internal energy and specific heat
for a simple cubic lattice $(z=6)$, are shown in Fig. \ref{fig2} , in comparison with DA \cite{refEFT} and EFT \cite{refEFTK1} as a limiting case  $c=1$ of
bond diluted system.

\begin{table}[h]\label{table1}
\begin{center}
\begin{threeparttable}
\caption{The critical temperatures of pure system with zero magnetic field obtained by several approximations as well as exact results and the results of the present work.}
\renewcommand{\arraystretch}{1.3}
\begin{tabular}{lllllllll}
\thickhline
Lattice & MFA & DA \cite{refACTA} &EFT\cite{refEFTK1} &  Present Work & Exact \cite{refexact}\cite{refexactz4}\\
\hline
$3$ & 3.0& 2.104 & 1.821 & 1.504& 1.519\\
$4$ & 4.0& 3.090 & 2.885 & 2.536& 2.269\\
$6$ & 6.0& 5.073 & 4.933 & 4.527& 4.511\\
$8$ & 8.0& 7.061 & 6.952 & 6.516& 6.353\\
$12$& 12.0& 11.045 & 10.970 & 10.499& 9.795\\
\thickhline \\
\end{tabular}
\end{threeparttable}
\end{center}
\end{table}


\subsection{Bond Diluted System}

Let us treat bond diluted system with zero magnetic field. The bond distribution function is given by
\eq{eq33}{P_J\paran{J_{ij}}=c\delta\paran{J_{ij}-J}+\paran{1-c}\delta\paran{J_{ij}}}
and it distribute bonds randomly between lattice sites to be $c$ percentage of bonds are closed and remaining $1-c$ percentage of bonds are open i.e.
$c$ is the concentration of closed bonds in the lattice.

Calculating the coefficients given in  Eqs. \re{eq15} and \re{eq21} by using Eq. \re{eq33} yields

\eq{eq34}{
\begin{array}{lcl}
A_n&=&\paran{\begin{array}{c}z\\n\end{array}} \left[c\cosh{\paran{J\nabla}}-c+1\right]^{z-n}c^n\sinh^{n}{\paran{J\nabla}}f(x)|_{x=0}\\
B_1&=&\left[c\cosh{\paran{J\nabla}}-c+1\right]f(x+\gamma)|_{x=0}\\
B_2&=&c\sinh{\paran{J\nabla}}f(x+\gamma)|_{x=0}
\end{array}
} By applying the procedure used between Eqs. \re{eq31} and \re{eq32} to Eq. \re{eq34} gives
\eq{eq35}{
\begin{array}{lcl}
A_n&=&\paran{\begin{array}{c}z\\n\end{array}}c^n\summ{p=0}{z-n}{}\summ{r=0}{p}{}\summ{s=0}{n}{}
\paran{\begin{array}{c}z-n\\p\end{array}}\paran{\begin{array}{c}p\\r\end{array}}
\paran{\begin{array}{c}n\\s\end{array}}\frac{c^p}{2^{n+p}}(1-c)^{z-n-p}\paran{-1}^{s}f\paran{n+p-2r-2s}\\
B_1&=&\frac{c}{2}\left[f(J+\gamma)+f(-J+\gamma)\right]+(1-c)f(\gamma)\\
B_2&=&\frac{c}{2}\left[f(J+\gamma)-f(-J+\gamma)\right]
\end{array}
}

Solving Eqs. \re{eq26} and \re{eq28} with coefficients given in Eqs. \re{eq35} and \re{eq27} gives the correlations and magnetization as a function of temperature.
Again, the critical temperature of the system for a given set of Hamiltonian parameters can be obtained by letting $\gamma\rightarrow 0$.
In this way the phase diagrams and the representative correlations of bond diluted s-1/2 Ising system with arbitrary coordination number $z$ can be obtained.

\begin{figure}[h]
\epsfig{file=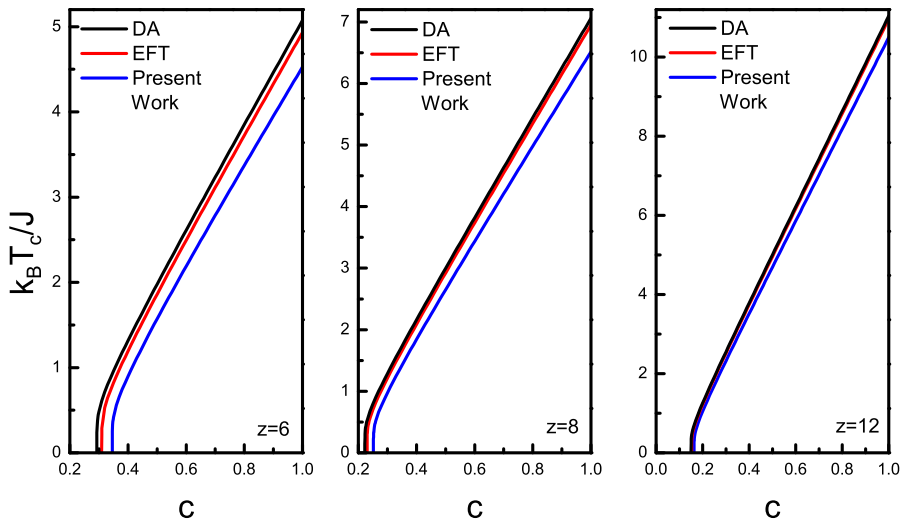, angle=0,width=12cm}
\caption{Variation of critical temperature  for  simple cubic lattice ($z=6$), body centered cubic lattice ($z=8$) and face centered cubic lattice ($z=12$)  with bond concentration.}\label{fig1}\end{figure}

The phase diagrams in $(k_BT_c/J-c)$ plane for three dimensional lattices
with coordination numbers $z=6$ (simple cubic lattice), $z=8$ (body centered cubic lattice)
and $z=12$ (face centered cubic lattice) can be seen in Fig. \ref{fig1}.
As seen in this figure, as the bond concentration decreases then the critical temperature values decrease gradually and fall to zero, as expected.
As seen in Fig. \ref{fig1} our formulation
gives lower values for critical temperatures at all concentration values with respect to other two approximations.
It is well known fact that for concentration values which are below a certain $c^{*}$, the system exhibits no ordered phase at all.
This specific $c^{*}$ value is called the critical bond concentration value. Since the formulation presented here gives lower critical
temperature values at all concentration values with respect to other EFT formulations, it will give also higher critical
bond concentration values. The critical bond concentration values of different lattices  can be seen in Table \ref{table2},
in comparison with the other two approximations.

\begin{figure}[h]
\epsfig{file=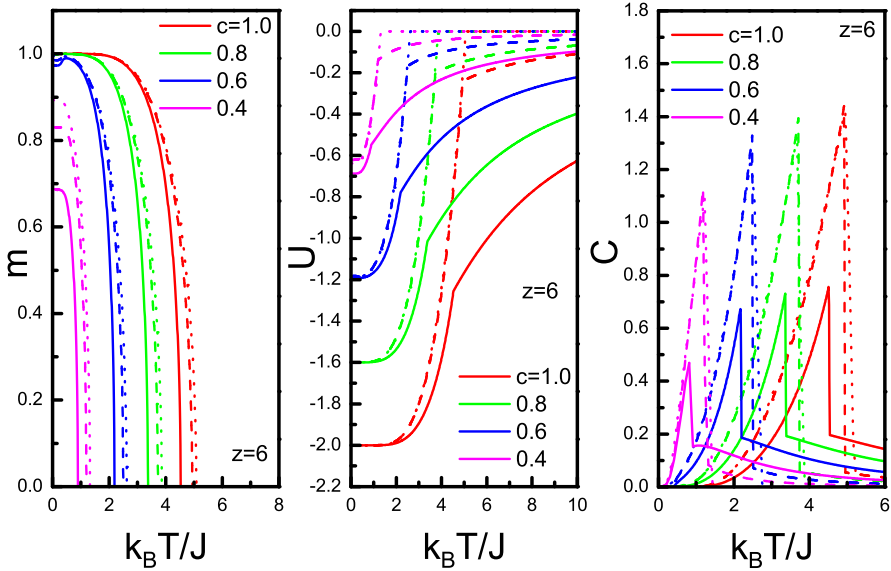, angle=0,width=12cm} \caption{Variation of
magnetization ($m$), internal energy ($U$) and specific heat ($C$)
per spin of simple cubic lattice with temperature for some selected
bond concentration values. The continuous lines are the results of
the presented formulation, dashed lines are the result of EFT and
the dotted lines are of DA.  }\label{fig2}\end{figure}

The internal energy and specific heat curves as function of Hamiltonian parameters and temperature can be easily obtained within the present formulation, as well as
better results for critical temperatures since the formulation is capable
of calculating the multi-site correlations as well as its single site counterparts in a representative manner. This can be seen in Fig. \ref{fig2} in comparison with DA and EFT. As
discussed above, since the DA neglects all multi-site correlations it will give zero internal energy
just above the critical temperature of the system which is physically impossible. On the other hand EFT gives more reasonable results than DA.

In Fig. \ref{fig2}, other than critical temperature values,  clear distinction stands out  between behaviors of internal energy just above the critical temperature.  EFT gives energy values more close to zero for the temperatures $T>T_c$ than our formulation. Also the difference of the behaviors of energy just above the critical temperature gives rise to a difference between specific heat behaviors after $T_c$ between EFT and formulation presented here.

\begin{figure}[h]
\epsfig{file=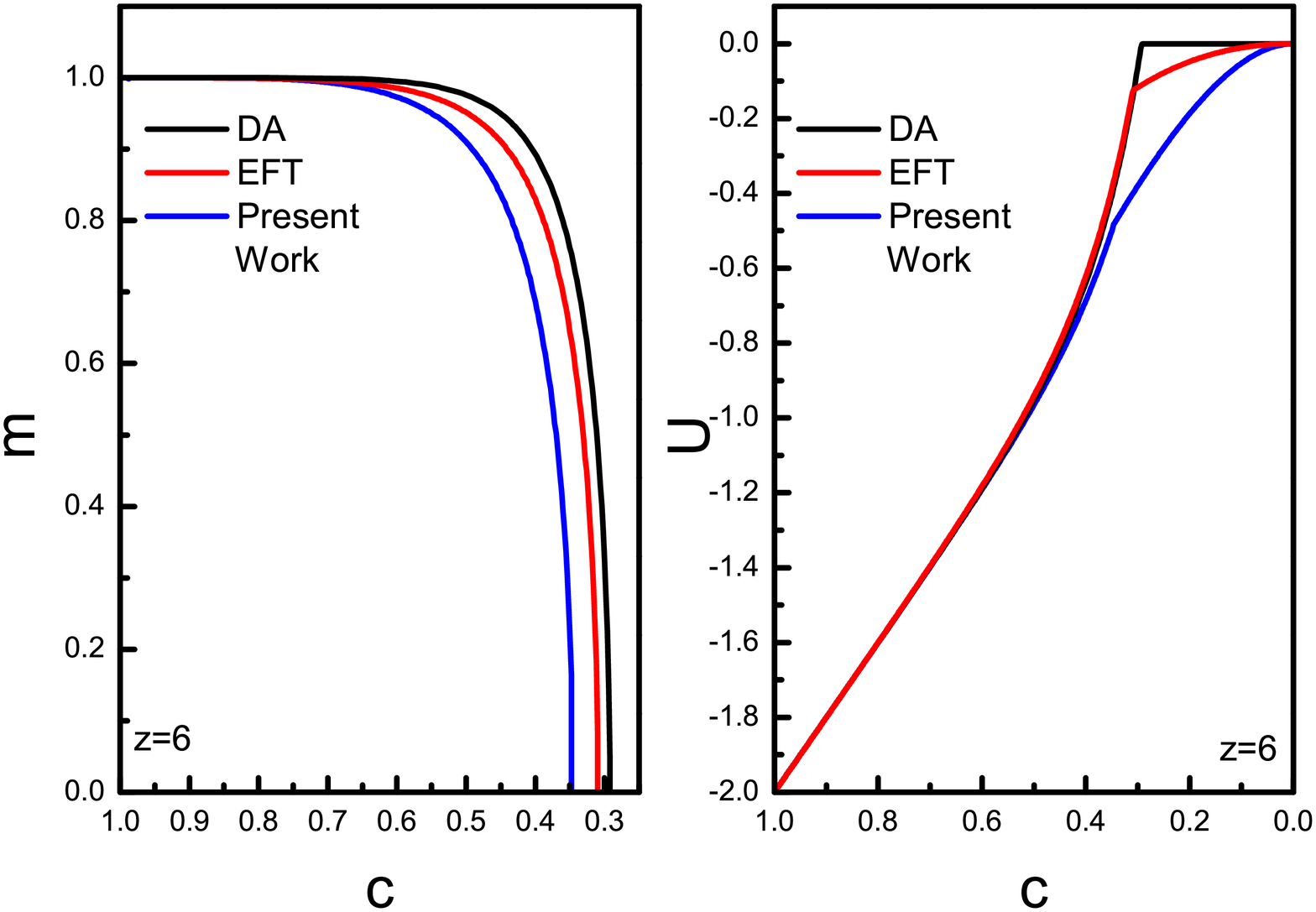, angle=0,width=12cm} \caption{Variation
of ground state magnetization and internal energy of simple cubic
lattice ($z=6$) with bond concentration. The calculations have been
done at the temperature $T=0.05$.}\label{fig4}\end{figure}

The other distinction shows itself in the ground state values of magnetization and  internal energy when  $c$ values get closer to the critical concentration value $c^{*}$.
For these concentration values, our
formalism gives lower values than EFT and DA for the ground state magnetization and internal energy. This result is expected since, while the $c$ starting to decrease
from $c=1$, the low temperature value of magnetization gradually decreases until $c$ reaches critical bond concentration value and
our formalism gives higher critical values for bond concentration .  The variation of the ground state
values of this thermodynamic functions can be seen in Fig. \ref{fig4} for simple cubic lattice ($z=6$).

As seen in Fig. \ref{fig4}, variation of the ground state values of magnetization and internal energy with bond concentration $c$ for simple cubic
lattice shows no significant difference for high $c$ values, i.e. where bond diluted system is close to pure system. In contrast to this,
the difference between our formulation and the results of EFT and DA for
for ground state values of magnetization and internal energy shows itself
for lower $c$ values.

\begin{table}[h]\label{table2}
\begin{center}
\begin{threeparttable}
\caption{The critical bond concentration values of bond diluted system with zero magnetic field obtained by
DA and EFT, as well as the results of the present work.
}
\renewcommand{\arraystretch}{1.3}
\begin{tabular}{llll}
\thickhline
Lattice & DA\cite{refACTA} & EFT\cite{refEFTK1} &  Present Work \\
\hline
$3$  & 0.5575 & 0.6623 & 0.7622\\
$4$  & 0.4284 & 0.4774 & 0.5470 \\
$6$  & 0.2929 & 0.3095 & 0.3458 \\
$8$  & 0.2224 & 0.2303 & 0.2520 \\
$12$  & 0.1504 & 0.1528 & 0.1633 \\
\thickhline \\
\end{tabular}
\end{threeparttable}
\end{center}
\end{table}

\section{Conclusion}\label{conclusion}

In this work, we present a general formulation of s-1/2 nearest neighbor Ising system with arbitrary coordination number $z$. The superiority of this formulation lies under its capability of calculating correlations in a representative manner and this advantage shows itself in the results of critical temperatures and the variation of thermodynamic functions with temperature such as the magnetization and the internal energy. Formulation covers some quenched disorder effects since derivation starts with Hamiltonian \re{eq1} which include bond disorder.

Disorder effects are important in material science since disorder (like bond dilution) induce important macroscopic effects in materials. Thus it is important to obtain critical values (e.g. critical temperatures, critical bond concentrations ) as well as variation of order parameter or some other thermodynamic functions (e.g. specific heat) with temperature as much as possible to the exact ones. It is a well known fact that it is impossible to obtain exact results for systems with disorder in most cases. On the other hand MC or similar simulation algorithms give accurate results for these systems but with some computational cost.

On the other hand, as mentioned in \cite{refdis1},\cite{refdis2} some diluted antiferromagnets in uniform external magnetic field corresponds to a ferromagnet in a random external magnetic field. Then in some cases, one can obtain
the behavior of more complex systems by solving s-1/2 Ising model or it's variants like random field distributed system.

This work is not the first attempt to handle the correlations which appears when expanding exact spin identities like \re{eq2}. Although the most of works deal with critical behavior of spin systems within the framework of EFT, these works are based on DA which means that neglecting all multi-site correlations, there are some works handling these correlations\cite{refEFT}, \cite{refEFTK1}. These works give results for critical temperatures as BPA. The importance of calculating these correlations are two fold. Firstly one can obtain more accurate critical values about the system and secondly one can obtain reasonable values for thermodynamic functions which are obtained from these correlations such as internal energy and the specific heat.

However, although this work is not the first attempt to handle correlations, the formulation presented here treat these correlations in a different way which give rise to more accurate results than small number of works related to it. Beyond that, we believe that it is important to obtain general formulation which covers arbitrary lattice and arbitrary Hamiltonian as long as it includes nearest neighbor interaction as a spin-spin  interaction. We hope that the formulation and
results obtained in this work may be beneficial form both theoretical and experimental point of view.

\end{document}